%% file: charm.tex
\documentclass[12pt]{article}
\usepackage{graphicx}

\newcommand\pubnumber{}
\newcommand\pubdate{\today}

\def\nara{Physics Department\\
Nara Women's University, \\
Kitauoya Nishi-machi, Nara, Japan -630 8506.}

\def\Journal#1#2#3#4{{#1} {\bf #2}, #3 (#4)}
  
\def\Title#1{\begin{center} {\Large #1 } \end{center}}
\def\Author#1{\begin{center}{ \sc #1} \end{center}}
\def\Address#1{\begin{center}{ \it #1} \end{center}}

\newcommand\pubblock{\rightline{\begin{tabular}{l} \pubnumber\\
         \pubdate  \end{tabular}}}
\newenvironment{Abstract}{\begin{quotation}  }{\end{quotation}}
\newenvironment{Presented}{\begin{quotation} \begin{center} 
             PRESENTED AT\end{center}\bigskip 
      \begin{center}\begin{large}}{\end{large}\end{center} \end{quotation}}
\def\Acknowledgements{\bigskip  \bigskip \begin{center} \begin{large}
             \bf ACKNOWLEDGEMENTS \end{large}\end{center}}

\def\PLB{{ Phys. Lett.}  B}
\def\PRL{ Phys. Rev. Lett.}
\def\PRD{{ Phys. Rev.} D}

\def\PR{{ Phys. Rep.}}

\def\Bpm{B^{\pm}}

\def\c1hi{\chi_{c1}}
\def\c2hi{\chi_{c2}}
\def\br{\mathcal{B}}
\def\1gk{B^{\pm} \to (\chi_{c1} \gamma) K^{\pm}}
\def\2gk{B^{\pm} \to (\chi_{c2} \gamma) K^{\pm}}

\def\kpsi2{B^{\pm} \to \psi_2(\to \chi_{c1} \gamma) K^{\pm}}

\input econfmacros.tex

\begin{document}
\begin{titlepage}
\pubblock

\vfill
\Title{Studies of radiative $X(3872)$ decays at Belle}
\vfill
\Author{ Vishal Bhardwaj (for the Belle Collaboration)}
\Address{\nara}
\vfill
\begin{Abstract}
We present a study of  the radiative decays of $X(3872)$ at Belle. In the
$\chi_{c1}\gamma$ final state, we got the first evidence of 
a new particle at 3823 MeV$/c^2$.
\end{Abstract}
\vfill
\begin{Presented}
The $5^{\rm th}$ International Workshop on Charm Physics\\
(Charm 2012)\\
14-17 May 2012, Honolulu, Hawai'i 96822.

\end{Presented}
\vfill
\end{titlepage}
\def\thefootnote{\fnsymbol{footnote}}
\setcounter{footnote}{0}

\section{Introduction}
 The Belle detector is  a general purpose spectrometer built to test the
Standard Model mechanism for $CP$-violation in $B$ decays to charmonium 
(golden channel)~\cite{cpv}. Parallel to this, Belle has proven 
to be an ideal place to carry out charmonium spectroscopy thanks to its 
very clean environment. Many new $c\bar{c}$ and $c\bar{c}$-like exotic 
candidates states such as $\eta_c(2S)$, $X(3872)$, $X(3915)$, 
$Z(3930)$, $X(3940$), $Z_1(4050)^+$, $Z_2(4250)^+$, $Y(4260)$, 
$Z(4430)^+$ and $Y(4660)$ have been found. 
Among them $X(3872)$ is the most interesting state which
was first observed in $B^\pm \to (J/\psi \pi^+ \pi^-) K^\pm$ at 
Belle~\cite{belle1}. Soon after its discovery, it was confirmed by the
CDF~\cite{cdf1}, D0~\cite{do1} and BaBar~\cite{babar0} collaborations. 
Recently, it has also been observed at  LHCb~\cite{lhcb} and CMS~\cite{cms}. 
The observation of $X(3872)$ in the same final states ($J/\psi \pi^+ \pi^-$) 
in  six different experiments,  reflects the eminent status of the $X(3872)$.
$X(3872)$'s narrow width and the proximity of its mass, $3871.7\pm0.2$ MeV$/c^2$
to the $D^{*0}\bar{D^0}$ threshold makes it a good candidate for a
$D\bar{D}^*$ molecule~\cite{swanson}. Other possibilities have also 
been proposed for the $X(3872)$ state, such as tetraquark~\cite{Maiani}, 
$c\bar{c}g$ hybrid meson~\cite{Li} and vector glueball models~\cite{Seth}. 

Radiative decays of $X(3872)$ provide  a unique opportunity to understand the
nature of $X(3872)$. For example, $X(3872) \to J/\psi \gamma$ resulted
in the confirmation of $C$-even ($C=+$) parity for 
$X(3872)$~\cite{belle2,babar1,belle3}. A 
similar decay mode $X(3872)\to \psi'\gamma$  can help 
to  identify $X(3872)$ as a charmonium, molecular or mixture of those states
\cite{swanson,mehen,wang}. Belle reported no significant signal~\cite{belle3}
and it  disagrees with the evidence at BaBar~\cite{babar1}. 

A recent search for the charged tetraquark  partner of $X(3872) 
(\to J/\psi \pi^+ \pi^0)$ gave negative results
\cite{belle_recent}. But still it is  hard to  totally rule out $X(3872)$ 
as a tetraquark, as some tetraquark models predict $X(3872)^+$ to be broad,
 thus still difficult to observe at current statistics~\cite{terasaki}. 
On the other hand, in both molecular and tetraquark hypothesis
a $C$-odd parity ($C=-$) partner can exist and it may dominantly decay  
into $\chi_{c1}\gamma$ and $\chi_{c2}\gamma$ final states. Partner searches would
be another approach to reveal $X(3872)$ structure. Along with this, undiscovered
$^3D_2$ charmonium ($\psi_2$) is expected to have a significant branching fraction
to $\chi_{c1}\gamma$ and $\chi_{c2}\gamma$~\cite{estia2002,cho1994}.
In this report, we also describe the first evidence of $\psi_2 \to \chi_{c1}
\gamma$ decay.

\section{Reconstruction}
$B$ mesons reconstructed using 
$B^{\pm} \to (\chi_{c1}(\to J/\psi \gamma) \gamma)  K^{\pm}$ and 
$B^{\pm} \to (\chi_{c2}(\to J/\psi \gamma) \gamma) K^{\pm}$
decay modes are used in the search for $C$-odd partner of the $X(3872)$ and 
other new narrow resonances. The results presented here are obtained 
from a data sample of $772\times 10^{6}$ $B\overline{B}$  events collected 
by the Belle 
detector~\cite{abashian} at the KEKB~\cite{kurokawa} 
energy-asymmetric $e^+e^-$ collider operating at the $\Upsilon(4S)$ resonance.

The $J/\psi$ meson is reconstructed in its decays to $\ell^+\ell^-$ 
($\ell =$ $e$ or $\mu$). 
In $e^+e^-$ decays, the four-momenta of all photons within 50 mrad of each 
of the original $e^+$ or $e^-$ tracks are included in the invariant mass 
calculation  $[$hereafter denoted as  $M_{e^+e^- (\gamma)}]$, in order 
to reduce the radiative tail. The reconstructed invariant mass of the $J/\psi$ 
candidates is required to satisfy $2.95 < M_{e^+ e^-(\gamma)} 
< 3.13$ GeV$/c^2$ or $3.03  < M_{\mu^+ \mu^-} < 3.13$ GeV$/c^2$.  
A mass- and vertex-constrained fit is applied to all the selected 
$J/\psi$ candidates in order to improve their momentum 
resolution.  The $\chi_{c1,c2}$ candidates are reconstructed 
by combining $J/\psi$ candidates with a photon having energy ($E_{\gamma}$) 
 greater than 200 MeV. The photons are reconstructed
from the energy deposition in electromagnetic calorimeter. 
To reduce the background from $\pi^0 \to
\gamma\gamma$, we calculate a likelihood function to distinguish an isolated
photon from $\pi^0$ decays using the photon pair's invariant mass, smaller 
photon's laboratory energy and polar angle~\cite{kopenberg}. Then we reject a
 photon having the $\pi^0$ likelihood ratio greater than 0.7 by combining 
with any other photon. The reconstructed invariant mass of
 $\chi_{c1}$  and $\chi_{c2}$ is required to satisfy 3.467 $< M_{J/\psi\gamma} <$ 
3.535 GeV$/c^2$ and 3.535 $< M_{J/\psi \gamma} <$ 3.611 GeV$/c^2$. A mass- and 
vertex-constrained fit is again performed to all the selected $\chi_{c1}$ and
 $\chi_{c2}$ candidates in order to improve their momentum resolution. 
Charged tracks are identified as  a kaon using information from the particle
identification devices. 

To reconstruct the $B$ candidates, each $\chi_{c1}$, $\chi_{c2}$ 
is combined with a kaon candidate and an additional photon having 
$E_{\gamma} > $ 100 MeV. In this additional photon selection, we reject the 
photon which in combination with another photon in that event, 
gives mass in the region around $\pi^0$ mass defined as 117 $< M_{\gamma\gamma}<$
153 MeV$/c^2$. In order to remove the reflection of photons coming from
 $\chi_{c1}$ and $\chi_{c2}$ decays, we 
reject the best $\chi_{c1}$ or $\chi_{c2}$ daughter photon from the additional
photon list to form a $B$ candidate. 
To identify the $B$ candidate, two kinematic variables are used : 
energy difference $\Delta E\equiv 
E_{B}^* - E_{beam}^*$ and beam-energy constrained mass $M_{\rm bc}\equiv 
\sqrt{(E_{beam}^*)^2 - (p_B^{cms})^2}$, where $E_{B}^*$ is the center-of-mass frame
 (cms) beam energy, and $E_{B}^*$ and $p_B^*$ are the  cms 
energy and momentum of 
the reconstructed particles. In case of multiple candidates, $\Delta E$ closest
to 0 is selected as the best one. Invariant mass of the final state 
($M_{\chi_{c1}\gamma}$ and $M_{\chi_{c2}\gamma}$) is used to identify the resonance.
In order to improve the resolution of $M_{\chi_{c\rm{x}} \gamma}$~\cite{CX}, we scale the energy of $\gamma$ to make $\Delta E$ equal to zero.

To suppress continuum background, events having a ratio of the second to zeroth 
Fox-Wolfram moments~\cite{foxwolfram} $R_2 > 0.5$ are rejected. 
Large $B\to J/\psi X$ MC samples (corresponding to $100$ times the data 
sample size used in this analysis) are used to study the background. 
The non-$J/\psi$ (non-$\chi_{c\rm{x}\gamma}$) background is studied using the 
$M_{\ell\ell}$  ($M_{J/\psi\gamma}$) sidebands in data. 
$\1gk$ and $\2gk$ yields are extracted from a 2D UML fit applied to 
the distribution in the $M_{\chi_{c \rm{x}} \gamma}$- $M_{\rm bc}$ space.

\begin{figure}[h!]
\begin{center}
  \includegraphics[angle=0,width=0.6\textwidth]{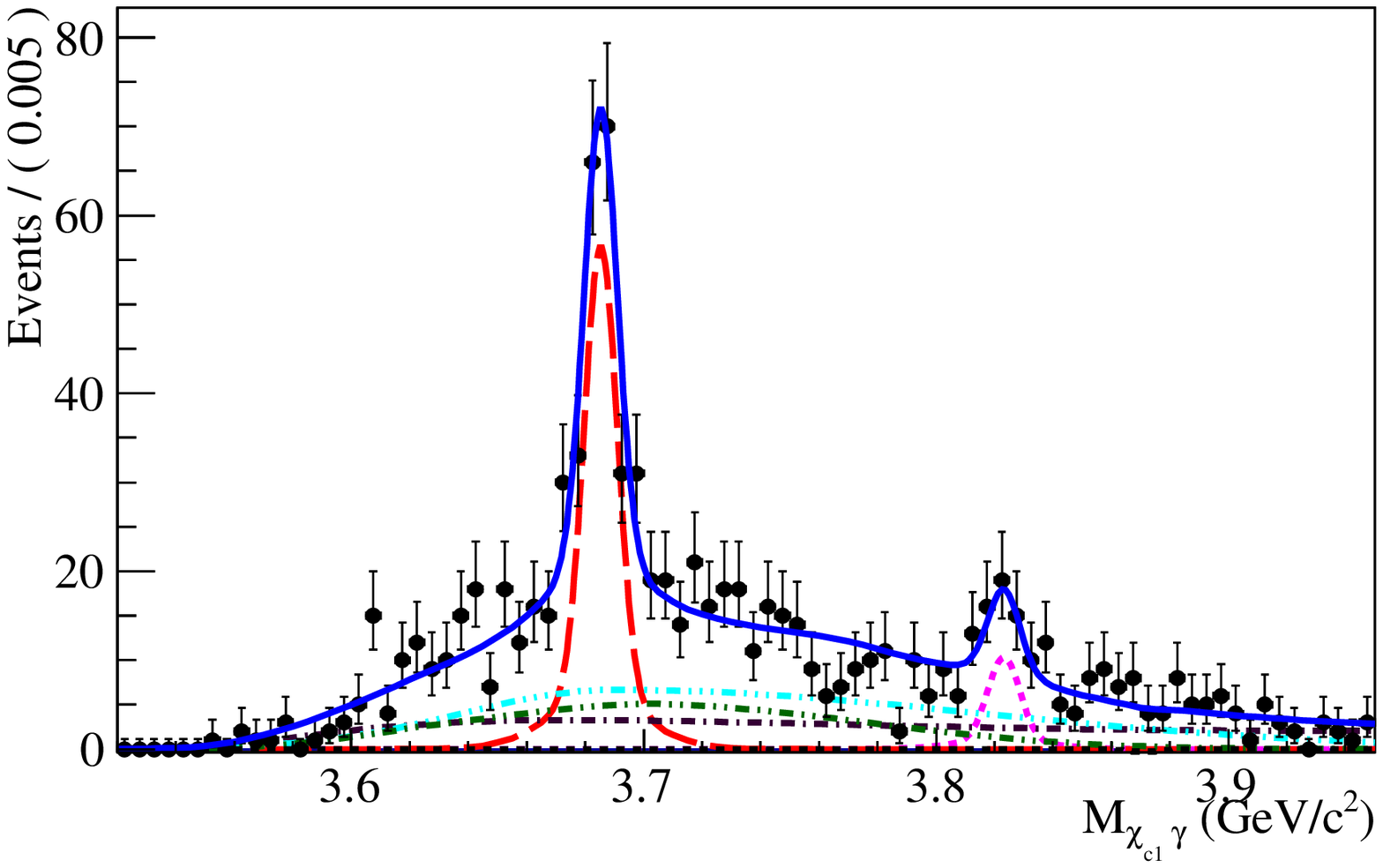} 
  \includegraphics[angle=0,width=0.6\textwidth]{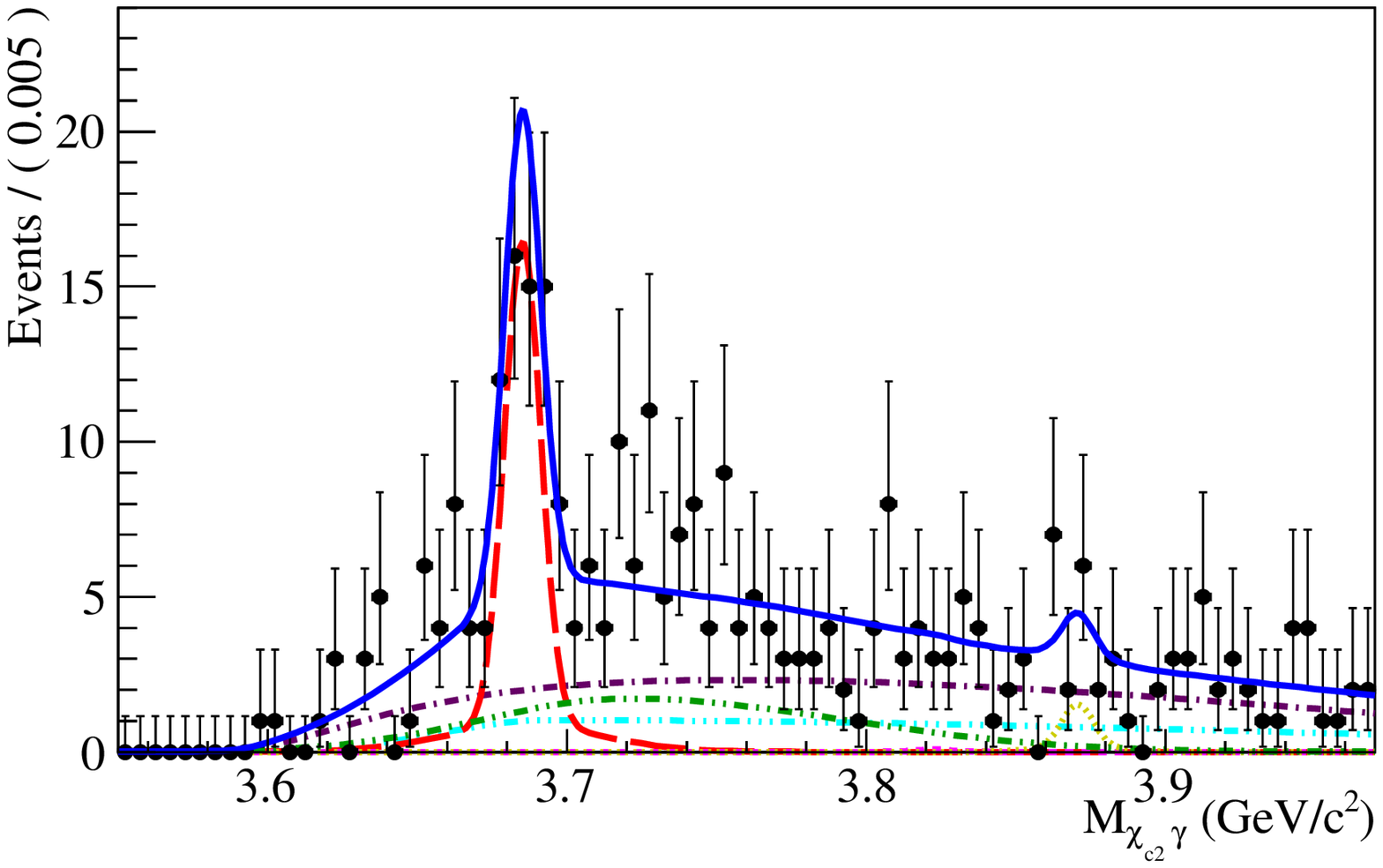}

  \caption{2D UML fit projections in $M_{\chi_{c1}\gamma}$ (top) and  $M_{\chi_{c2}\gamma}$ (bottom) for the signal region $M_{\rm bc} > 5.27$ GeV$/c^2$.
The curves show the signal [red large-dashed for $\psi'$, pink dashed for $\psi_2$ and yellow dotted for $X(3872)$] and the background component [purple dotted-dashed for combinatorial, dark green two dotted-dashed for $B^{\pm}\to\psi' ($other than $\chi_{c1}\gamma)K^{\pm}$ and cyan three dotted-dashed for peaking component] as well as the overall fit [blue solid].  }
\label{fig:both}
\end{center}
\end{figure}

\section{Results}
In $\1gk$ decay, a clear peak of $\psi' \to \chi_{c1}\gamma$ is observed
 in the $\chi_{c1}\gamma$ invariant mass ($M_{\chi_{c1}\gamma}$) projection.
In addition to  this, we  also find  clear evidence of a narrow peak at 
3823 MeV$/c^2$, denoted as $X(3820)$ hereafter.
Since statistics are still limited, its width is poorly constrained to be
$4\pm6$ MeV even if it is floated in the fit. Therefore we set $X(3820)$
natural width to be 0 MeV for further discussion.
 We estimate the statistical significance of 
$X(3820) \to\chi_{c1}\gamma$ to be  4.2$\sigma$ including systematic uncertainty.
 In our search for $X(3872) \to \chi_{c1}\gamma$, no signal is seen 
 and 90\% confidence level (C.L.) upper limit (U.L.) on
$\br(\Bpm \to X(3872)K^{\pm}) \br(X(3872) \to\chi_{c1}\gamma)$ is estimated
using a frequentist approach. In $\2gk$, we observe a clear peak of the
$\psi' \to \chi_{c2}\gamma$. However, we do not see any hint of a narrow 
resonance with the current statistics. Figure  \ref{fig:both} shows the 
projections of  2D UML fits to the $M_{\chi_{c1}\gamma}$ and $M_{\chi_{c2}\gamma}$ cases
for the signal region, $M_{\rm bc} > 5.27$ GeV$/c^2$. 
Table \ref{tab:br} summarizes the results of $\1gk$ and $\2gk$ decays.

The mass of $X(3820)$ is near the potential model expectations
for the center-of-gravity (cog) of $1^3D_J$ states: the Cornell~\cite{cornell} 
and the Buchm\"{u}ller-Tye~\cite{buchmuller} potential, which gives 
$M_{\rm cog}$ (1D) = 3810 MeV$/c^2$. Some models predict the mass of 
$^3D_2$ ($J^{PC}= 2^{--}$)  states to be 3815-3840 MeV$/c^2$~\cite{Godfrey,Eichten}. $X(3820)$ mass agrees quite well within the expectation. 
The $X(3820)$ state is likely to be the missing $ ^3D_2~ c\bar{c}$ ($\psi_2$) state.

\begin{table}[t]
\vspace{0.4cm}
\begin{center}
\begin{tabular}{|c|c|c|}
\hline 
 Channel & Yield  & $\mathcal{B} (10^{-4})$ \\ \hline

$B^{\pm}\to \psi'(\to \chi_{c1} \gamma) K^{\pm}$ & $193.2^{+19.2}_{-18.6}$    & $7.7^{+0.8+0.9}_{-0.7-0.8}$ \\ 
$B^{\pm}\to \psi'(\to \chi_{c2} \gamma) K^{\pm}$ & $59.1^{+8.4}_{-8.0}$ & $6.3 \pm0.9\pm0.6$ \\ 

\hline
& & $\times 10^{-6}$  \\ \hline
$B^{\pm}\to X(3820)(\to \chi_{c1}\gamma) K^{\pm}$  & $33.2^{+9.2}_{-8.5}$ & $9.7^{+2.8+1.1}_{-2.5-1.0}$ \\
$B^{\pm}\to X(3872)(\to \chi_{c1}\gamma) K^{\pm}$ & $-0.9 \pm 5.1$ & $< 2.0$  \\ 
$B^{\pm}\to X(3820)(\to \chi_{c2}\gamma) K^{\pm}$  & $0.3^{+3.9}_{-3.1}$ & $<3.6$ \\ 
$B^{\pm}\to X(3872)(\to \chi_{c2}\gamma) K^{\pm}$ & $4.7^{+4.4}_{-3.6}$ & $< 6.7$  \\ \hline

\end{tabular}
\caption{ Summary of the results. Measured $\br$ (with 90\% confidence level (C.L.) upper limit (U.L.) for $B^{\pm}\to X(3872)(\to \chi_{c1}\gamma) K^{\pm}$,
$B^{\pm}\to X(3823)(\to \chi_{c2}\gamma) K^{\pm}$ and 
$B^{\pm}\to X(3872)(\to \chi_{c2}\gamma) K^{\pm}$ decay modes).  $\mathcal{B}$ 
for $B\to X K$ is 
$\mathcal{B} (B\to X K)  \mathcal{B} (X \to \chi_{c\rm x} \gamma)$, here $X$ stands for $X(3820)$ and $X(3872)$. For $\br$, the first (second) error is statistical (systematic). }

\label{tab:br}
\end{center}
\end{table}

\section{Summary}
In the study of  $\1gk$ and $\2gk$ decays, Belle observe the $\psi'$ peak
in both decay modes and other peak is found to be consistent with 
expectation. In $\1gk$, Belle find the first evidence of
a  narrow state having mass of $3823.5 \pm 2.5$ MeV$/c^2$. 
This narrow state is likely to be the missing $\psi_2$ ($^3D_2$ $c\bar{c}$ 
state) because the observed mass totally agrees with a theoretical expectation
and $\chi_{c1}\gamma$ is one of the dominant decay modes.
While in the search of $C$-odd partner of $X(3872)$, no signal is found at the
current statistics  and U.L. in 90\% C.L. are provided.

\Acknowledgements
Authors's participation to the 5th International
Workshop on Charm Physics (Charm2012) was supported
by MEXT KAKENHI, Grant-in-Aid for Scientific Research
on Innovative Areas, entitled
“Elucidation of New hadrons with a Variety of Flavors”.

\end{document}

%% file: econfmacros.tex
\def\beq{\begin{equation}}
\def\eeq#1{\label{#1}\end{equation}}
\def\eeqn{\end{equation}}

\def\beqa{\begin{eqnarray}}
\def\eeqa#1{\label{#1}\end{eqnarray}}
\def\eeqan{\end{eqnarray}}

\let\bar=\overbar

\def\Dslash{\not{\hbox{\kern-4pt $D$}}}
\def\dslash{\not{\hbox{\kern-2pt $\del$}}}

\def\msb{{\bar{\ssstyle M \kern -1pt S}}}

